\documentclass[letterpaper, 10 pt, conference]{ieeeconf}  

\IEEEoverridecommandlockouts                              

\usepackage{amsfonts}
\usepackage{mathrsfs}
\usepackage{amssymb}
\usepackage{amsmath}
\usepackage{mathrsfs}
\usepackage{array}
\usepackage{extarrows}
\usepackage{color}
\usepackage{cite}
\usepackage{graphicx}
\usepackage[font=footnotesize]{caption}
\usepackage[font=footnotesize]{subcaption}
\DeclareMathOperator{\grank}{grank \,}

\DeclareMathOperator{\col}{col \,}
\DeclareMathOperator{\row}{row \,}

\makeatletter 
\newsavebox\myboxA
\newsavebox\myboxB
\newlength\mylenA
\newcommand*\xoverline[2][0.86]{%
    \sbox{\myboxA}{$\m@th#2$}%
    \setbox\myboxB\null
    \ht\myboxB=\ht\myboxA%
    \dp\myboxB=\dp\myboxA%
    \wd\myboxB=#1\wd\myboxA
    \sbox\myboxB{$\m@th\overline{\copy\myboxB}$}
    \setlength\mylenA{\the\wd\myboxA}
    \addtolength\mylenA{-\the\wd\myboxB}%
    \ifdim\wd\myboxB<\wd\myboxA%
       \rlap{\hskip 0.5\mylenA\usebox\myboxB}{\usebox\myboxA}%
    \else
        \hskip -0.5\mylenA\rlap{\usebox\myboxA}{\hskip 0.5\mylenA\usebox\myboxB}%
    \fi}

\makeatletter
\newcommand*\bigcdot{\mathpalette\bigcdot@{.5}}
\newcommand*\bigcdot@[2]{\mathbin{\vcenter{\hbox{\scalebox{#2}{$\m@th#1\bullet$}}}}}
\makeatother

\input{steve.lib}

\title{\LARGE \bf
On the Existence of a Fixed Spectrum for a Multi-channel Linear System: A Matroid Theory Approach
}

\author{F. Liu$^{1}$ and A. S. Morse$^{1}$
\thanks{*This work was supported by National Science Foundation grant n. 1607101.00 and US Air Force grant n. FA9550-16-1-0290.}
\thanks{$^{1}$ F. Liu and A. S. Morse are with the Department
of Electrical Engineering, Yale University, New Haven, CT, USA.
        {\tt\small \{fengjiao.liu,  as.morse\}@yale.edu }}%
        }

\begin{document}
\maketitle 
\thispagestyle{empty} 
\pagestyle{empty}

\begin{abstract}
Conditions for the existence of a fixed spectrum \{i.e., the set of fixed modes\} for a multi-channel linear system have been known for a long time. The aim of this paper is to reestablish one of these conditions using a new and transparent approach based on matroid theory. 
\end{abstract}

\section{Introduction}

The classical ``decentralized control'' problem considered in \cite{wang1973stabilization, corfmat1976decentralized} focuses on stabilizing or otherwise controlling a $k>1$ channel linear system of the form
\begin{equation} \label{eqn:syst}
\dot{x} = Ax + \sum_{i=1}^k B_i u_i,\hspace{.3in} y_i = C_i x_i
\end{equation}
Decentralization is enforced by restricting the feedback of each measured signal $y_i$ to only its corresponding control input $u_i$, possibly through a linear dynamic controller. Wang and Davison \cite{wang1973stabilization} were able to show that no matter what these feedback controllers might be, as long as they are finite dimensional and linear time-invariant (LTI), the spectrum of the resulting closed-loop system contains a fixed subset depending only on $A$, the $B_i$ and the $C_i$, which they elected to called the set of ``fixed modes'' of the system. Roughly speaking, the set of fixed modes of (\ref{eqn:syst}), henceforth called the ``fixed spectrum'' of (\ref{eqn:syst}), is the the spectrum of $A$ that cannot be shifted by the decentralized output feedback laws $u_i = F_i y_i$, $i \in \{1, 2, \dots, k\}$. That is, if $A \in \mathbb{R}^{n \times n}$, $B_i \in \mathbb{R}^{n \times m_i}$, $C_i \in \mathbb{R}^{l_i \times n}$, the \emph{fixed spectrum} of (\ref{eqn:syst}), written $\Lambda_{\rm fixed}$, is precisely
\begin{equation*}
\Lambda_{\text{fixed}} = \bigcap_{F_i \in \mathbb{R}^{m_i \times l_i}} \Lambda \left( A+\sum_{i=1}^k B_i F_i C_i \right)
\end{equation*}
where $\Lambda(\cdot)$ denotes the spectrum. Since the $F_i$ can be zero, it is clear that the fixed spectrum is a subset of the spectrum of $A$. It is possible that the fixed spectrum is an empty set, in which case it is said that the system has no fixed spectrum.

Wang and Davison showed that $\Lambda_{\text{fixed}}$ is contained in the closed-loop spectrum of the system which results when any given finite dimensional LTI decentralized controls are applied to (\ref{eqn:syst}). Thus $\Lambda_{\text{fixed}}$ must be a stable spectrum if decentralized stabilization is to be achieved. Wang and Davison were also able to show that the stability of $\Lambda_{\text{fixed}}$ is sufficient for decentralized stabilization with linear dynamic controllers. Not surprisingly, the notion of a fixed spectrum also arises in connection with the decentralized spectrum assignment problem treated in \cite{corfmat1976decentralized, vidyasagar1982algebraic}. In particular it is known that a necessary and sufficient condition for ``free" assignability of a closed-loop spectrum with finite dimensional LTI decentralized controllers is that there is no fixed spectrum \cite{wang1973stabilization, corfmat1976decentralized}. However, it should be noted that unlike the centralized case, free spectrum assignability in the decentralized case presumes that the overall spectrum admits a suitable partition into a finite number of symmetric sets, the partition being determined by the strongly connected components in a suitably defined graph of (\ref{eqn:syst}) \cite{corfmat1976decentralized}. It is clear from the preceding that $\Lambda_{\text{fixed}}$ plays a central role in both the decentralized stabilization and decentralized spectrum assignment problems. Accordingly many characterizations of $\Lambda_{\text{fixed}}$ exist \cite{gong1992characterization}. A certain subset of a fixed spectrum can be eliminated by time-varying decentralized controllers \cite{anderson1981time, willems1989time}, sampling strategies \cite{ozguner1985sampling, willems1988elimination}, or other techniques \cite{lavaei2006elimination, stankovic2008stabilization}. The subset of a fixed spectrum which cannot be eliminated by any (including nonlinear) decentralized controllers are characterized in \cite{gong1997stabilization}. An explicit necessary and sufficient matrix-algebraic condition for a complex number $\lambda$ to be in the fixed spectrum of (\ref{eqn:syst}) is derived in \cite{anderson1981algebraic}, using matrix pencils and matrix nets \cite{anderson1982structural}. An equivalent algebraic condition is established in \cite{gong1994equivalence, liu1996comments}. Equivalent graph-theoretic criteria for the existence of a fixed spectrum are developed in \cite{reinschke1984graph}. Frequency domain characterizations of a fixed spectrum are presented in \cite{seraji1982fixed, anderson1982transfer, vidyasagar1983algebraic, xie1986frequency, xu1986further, kong1996graph}. In particular, the characterization of a fixed spectrum in \cite{anderson1981algebraic} is quite fundamental and inspires a number of later work \cite{gong1992characterization, gong1994equivalence, liu1996comments, gong1997stabilization, reinschke1984graph}.

Let $\mathbf{k} \triangleq \{1, 2, \dots, k\}$. Suppose $\mathcal{S} = \{i_1, i_2, \dots, i_s\} \subset \mathbf{k}$ with $i_1 < i_2 < \dots < i_s$, the complement of $\mathcal{S}$ in $\mathbf{k}$ is denoted by $\mathbf{k} - \mathcal{S} = \{j_1, j_2, \dots, j_{k-s}\}$ with $j_1 < j_2 < \dots < j_{k-s}$. Let 
\begin{equation*} 
B_\mathcal{S} \triangleq [B_{i_1} \enspace B_{i_2} \enspace \dots \enspace B_{i_s}], \qquad
C_{\mathbf{k} - \mathcal{S}} \triangleq 
\begin{bmatrix}
C_{j_1} \\
C_{j_2} \\
\vdots \\
C_{j_{k-s}}
\end{bmatrix}
\end{equation*}
The result in \cite{anderson1981algebraic} is stated below.

\begin{proposition} \textnormal{\cite{anderson1981algebraic}}  \label{prp:mtrx-pencil}
A $k$-channel linear system $\{A, B_i, C_i; k\}$ has $\lambda \in \Lambda(A)$ in its fixed spectrum if and only if $\exists \mathcal{S} \subset \mathbf{k}$ such that
\begin{equation} \label{eqn:real-partition}
\rank 
\begin{bmatrix}
\lambda I_n -A & B_{\mathcal{S}} \\
C_{\mathbf{k}-\mathcal{S}} & \mathbf{0}
\end{bmatrix}
< n
\end{equation}
\end{proposition}

Although originally proved with techniques like matrix pencils and matrix nets \cite{anderson1981algebraic, anderson1982structural}, Proposition \ref{prp:mtrx-pencil} reveals an insight that whether a multi-channel linear system has a fixed spectrum is indeed a combinatorial problem. By exploiting this fact, a new and transparent proof of Proposition \ref{prp:mtrx-pencil} using matroid theory is presented in this paper.

Roughly speaking, a matoid is a mathematical structure that generalizes the notion of linear independence in vector spaces. It is a handy tool for studying combinatorial properties of matrices.

\section{Proof}

A formal proof of Proposition \ref{prp:mtrx-pencil} is given in this section, for which some concepts and three lemmas are needed.

\begin{lemma} \label{lem:pencil-rank}
Let matrices $A \in \mathbb{C}^{n \times n}$, $B \in \mathbb{C}^{n \times m}$, and $C \in \mathbb{C}^{l \times n}$. Then $
\rank 
\begin{bmatrix}
A & B \\
C & \mathbf{0}
\end{bmatrix}
< n
$
if and only if $\rank (A + BE + KC) < n$ for any matrices $E \in \mathbb{C}^{m \times n}$ and $K \in \mathbb{C}^{n \times l}$.
\end{lemma}

\noindent \textbf{Proof of Lemma \ref{lem:pencil-rank}:} (Necessity) If $
\rank 
\begin{bmatrix}
A & B \\
C & \mathbf{0}
\end{bmatrix}
< n
$, then $
\rank 
\begin{bmatrix}
A+BE+KC & B \\
C & \mathbf{0}
\end{bmatrix}
< n
$
for any matrices $E \in \mathbb{C}^{m \times n}$ and $K \in \mathbb{C}^{n \times l}$, as the rank of a matrix remains unchanged under elementary row and column operations. So $\rank (A + BE + KC) < n$ for any $E$ and $K$.

(Sufficiency) Suppose $
\rank 
\begin{bmatrix}
A & B \\
C & \mathbf{0}
\end{bmatrix}
\geq n
$, by elementary column operations, $\exists E \in \mathbb{C}^{m \times n}$ such that $
\rank 
\begin{bmatrix}
A + BE \\
C 
\end{bmatrix}
= n
$. Similarly, by elementary row operations, $\exists K \in \mathbb{C}^{n \times l}$ such that $\rank (A + BE + KC) = n$. This completes the proof of Lemma \ref{lem:pencil-rank}. \hfill $\qed$

Let $M$ be a parameterized matrix whose entries are linear combinations of algebraically independent scalar parameters. The \emph{generic rank} of $M$, denoted by $\grank M$, is the maximum rank of $M$ that can be achieved as the parameters vary over the entire parameter space. It is generic in the sense that it is achievable by any parameter values in the complement of a proper algebraic set in the parameter space.

For some $d \in \mathbb{N}$, we are given column vectors $w_1$, $w_2$, $\dots$, $w_d \in \mathbb{C}^{n_1}$ and row vectors $r_1$, $r_2$, $\dots$, $r_d \in \mathbb{C}^{1 \times n_2}$. Let $\mathbf{d} \triangleq \{1, 2, \dots, d\}$. Suppose $\mathcal{S} = \{i_1, i_2, \dots, i_s\} \subset \mathbf{d}$ with $i_1 < i_2 < \dots < i_s$, and $\mathbf{d} - \mathcal{S} = \{j_1, j_2, \dots, j_{d-s}\}$ with $j_1 < j_2 < \dots < j_{d-s}$, let

\begin{equation} \label{eqn:agg-mtrx}
w_{\mathcal{S}} = [w_{i_1} \enspace w_{i_2} \enspace \dots \enspace w_{i_s}], \qquad
r_{\mathbf{d} - \mathcal{S}} = 
\begin{bmatrix}
r_{j_1} \\
r_{j_2} \\
\vdots \\
r_{j_{d-s}}
\end{bmatrix}
\end{equation}
Let $\mathcal{P} = \{(w_i, r_i) \, | \, i \in \mathbf{d}\}$ be the set of $d$ pairs of vectors. For $\mathcal{I} \subset \mathcal{P}$, let $\{w_i \, | \, (w_i, r_i) \in \mathcal{I}\}$ be the set of column vectors appearing in $\mathcal{I}$; let $\{r_i \, | \, (w_i, r_i) \in \mathcal{I}\}$ be the set of row vectors appearing in $\mathcal{I}$. Note that 
\begin{equation*}
\big|\mathcal{I}\big| = \big|\{w_i \, | \, (w_i, r_i) \in \mathcal{I}\}\big| = \big|\{r_i \, | \, (w_i, r_i) \in \mathcal{I}\}\big|
\end{equation*} 
where $|\mathcal{I}|$ is the number of vector pairs in $\mathcal{I}$. A nonempty subset $\mathcal{I} \subset \mathcal{P}$ is \emph{jointly independent} if $\{w_i \, | \, (w_i, r_i) \in \mathcal{I}\}$ and $\{r_i \, | \, (w_i, r_i) \in \mathcal{I}\}$ are both linearly independent sets. Let $\mathcal{J}\{w_i, r_i; d\}$ be the set of all jointly independent subsets of $\mathcal{P}$. The following proposition summarizes Lemma 1 (derived from the matroid intersection theorem \cite{murota2010matroids}) and Lemma 2 in \cite{liu2019graphical}.




\begin{proposition} \textnormal{\cite{liu2019graphical}}  \label{prp:grk}
For some $d \in \mathbb{N}$, given column vectors $w_1$, $w_2$, $\dots$, $w_d \in \mathbb{C}^{n_1}$, row vectors $r_1$, $r_2$, $\dots$, $r_d \in \mathbb{C}^{1 \times n_2}$, and algebraically independent scalar parameters $p_1$, $p_2$, $\dots$, $p_d$, the following equation holds.
\begin{align*}
\grank \left( \sum_{i \in \mathbf{d}} w_i p_i r_i \right) 
&= \max_{\mathcal{I} \in \mathcal{J}\{w_i, r_i; d\}} |\mathcal{I}| \\
&= \min_{\mathcal{S} \subset \mathbf{d}} \left( \rank w_{\mathcal{S}} + \rank r_{\mathbf{d}-\mathcal{S}} \right)
\end{align*}
\end{proposition}

Although Proposition \ref{prp:grk} was initially developed for real vectors and parameters, the same proof applies to complex vectors and parameters without change. Therefore a proof of Proposition \ref{prp:grk} will not be given here. The next lemma extends Proposition \ref{prp:grk} to a more general case.

For some $d \in \mathbb{N}$, we are given complex matrices $W_1$, $W_2$, $\dots$, $W_d$, each of which has $n_1$ rows and an arbitrary number of columns, and complex matrices $R_1$, $R_2$, $\dots$, $R_d$, each of which has an arbitrary number of rows and $n_2$ columns. Suppose $\mathcal{S} = \{i_1, i_2, \dots, i_s\} \subset \mathbf{d}$ with $i_1 < i_2 < \dots < i_s$, and $\mathbf{d} - \mathcal{S} = \{j_1, j_2, \dots, j_{d-s}\}$ with $j_1 < j_2 < \dots < j_{d-s}$, let 
\begin{equation*}
W_\mathcal{S} = [W_{i_1} \enspace W_{i_2} \enspace \dots \enspace W_{i_s}], \qquad
R_{\mathbf{d} - \mathcal{S}} = 
\begin{bmatrix}
R_{j_1} \\
R_{j_2} \\
\vdots \\
R_{j_{d-s}}
\end{bmatrix}
\end{equation*} 
For each $i \in \mathbf{d}$, let $\col(W_i)$ denote the set of column vectors of $W_i$ and let $\row(R_i)$ denote the set of row vectors of $R_i$. Let $\mathcal{P} = \{(w, r) \, | \, w \in \col(W_i), r \in \row(R_i), i \in \mathbf{d}\}$ be the set of all vector pairs taken from matrices with the same indices. For $\mathcal{I} \subset \mathcal{P}$, let $\{w \, | \, (w, r) \in \mathcal{I}, r \in \row(R_{\mathbf{d}})\}$ be the set of column vectors appearing in $\mathcal{I}$, with any common vectors repeated; let $\{r \, | \, (w, r) \in \mathcal{I}, w \in \col(W_{\mathbf{d}})\}$ be the set of row vectors appearing in $\mathcal{I}$, with any common vectors repeated. That is, if a column vector $w$ appears exactly $n_w$ times in $\mathcal{I}$, the set $\{w \, | \, (w, r) \in \mathcal{I}, r \in \row(R_{\mathbf{d}})\}$ contains $n_w$ copies of $w$. Note that 
\begin{align*}
\big|\mathcal{I}\big|
&= \big|\{w \, | \, (w, r) \in \mathcal{I},~ r \in \row(R_{\mathbf{d}})\}\big| \\
&= \big|\{r \, | \, (w, r) \in \mathcal{I},~ w \in \col(W_{\mathbf{d}})\}\big|
\end{align*}
A nonempty subset $\mathcal{I} \subset \mathcal{P}$ is \emph{jointly independent} if $\{w \, | \, (w, r) \in \mathcal{I},~ r \in \row(R_{\mathbf{d}})\}$ and $\{r \, | \, (w, r) \in \mathcal{I},~ w \in \col(W_{\mathbf{d}})\}$ are both linearly independent sets. Let $\mathcal{J}\{W_i, R_i; d\}$ be the set of all jointly independent subsets of $\mathcal{P}$.

\begin{lemma} \label{lem:grk-mtrx}
For some $d \in \mathbb{N}$, given complex matrices $W_1$, $W_2$, $\dots$, $W_d$, each of which has $n_1$ rows and an arbitrary number of columns, complex matrices $R_1$, $R_2$, $\dots$, $R_d$, each of which has an arbitrary number of rows and $n_2$ columns, and parameterized matrices $P_1$, $P_2$, $\dots$, $P_d$ of appropriate sizes, whose entries are algebraically independent thus are modeled by distinct parameters, the following equation holds.
\begin{align*}
\grank \left( \sum_{i \in \mathbf{d}} W_i P_i R_i \right) 
&= \max_{\mathcal{I} \in \mathcal{J}\{W_i, R_i; d\}} |\mathcal{I}| \\
&= \min_{\mathcal{S} \subset \mathbf{d}} \left( \rank W_{\mathcal{S}} + \rank R_{\mathbf{d} - \mathcal{S}} \right)
\end{align*}
\end{lemma}

As Proposition \ref{prp:grk} is proved directly from the matroid intersection theorem, it is natural to ask whether Lemma \ref{lem:grk-mtrx} can be proved in the same way. Unfortunately, the answer is no. In order to do that, we may need a more general notion of matroid which allows the rank of an element to be greater than 1.

\noindent \textbf{Proof of Lemma \ref{lem:grk-mtrx}:} For each $t \in \mathbf{d}$, suppose $W_t$ is of size $n_1 \times \alpha_t$, $R_t$ is of size $\beta_t \times n_2$, and $P_t$ is of size $\alpha_t \times \beta_t$. Note that $W_t P_t R_t = \sum\limits_{i=1}^{\alpha_t} \sum\limits_{j=1}^{\beta_t} w_t^i p_t^{ij} r_t^j$, where $w_t^i \in \mathbb{C}^{n_1}$ is the $i$th column of $W_t$, $p_t^{ij}$ is the $ij$th entry of $P_t$, and $r_t^j \in \mathbb{C}^{1 \times n_2}$ is the $j$th row of $R_t$. Recall that $\mathcal{P} = \{(w, r) \, | \, w \in \col(W_t), r \in \row(R_t), t \in \mathbf{d}\}$, now let $\bar{d} = |\mathcal{P}| = \sum\limits_{t \in \mathbf{d}} \alpha_t \beta_t$. Let the entries of the $P_t$, $t \in \mathbf{d}$, be denoted by $p_1$ through $p_{\bar{d}}$ and let the vector pairs in $\mathcal{P}$ be labeled $1$ through $\bar{d}$, i.e., $\mathcal{P} = \{(w_i, r_i) \, | \, i \in \bar{\mathbf{d}}\}$, where $\bar{\mathbf{d}} \triangleq \{1, 2, \dots, \bar{d}\}$. So $\mathcal{J}\{w_i, r_i; \bar{d}\} = \mathcal{J}\{W_i, R_i; d\}$. Thus 
\begin{equation*}
\sum_{t \in \mathbf{d}} W_t P_t R_t
= \sum_{t=1}^d \sum\limits_{i=1}^{\alpha_t} \sum\limits_{j=1}^{\beta_t} w_t^i p_t^{ij} r_t^j
= \sum_{i \in \bar{\mathbf{d}}} w_i p_i r_i
\end{equation*}
By Proposition \ref{prp:grk},
\begin{align*}
\grank \left( \sum_{i \in \mathbf{d}} W_i P_i R_i \right) 
&= \grank \left( \sum_{i \in \bar{\mathbf{d}}} w_i p_i r_i \right) \\
&= \max_{\mathcal{I} \in \mathcal{J}\{w_i, r_i; \bar{d}\}} |\mathcal{I}| \\
&= \max_{\mathcal{I} \in \mathcal{J}\{W_i, R_i; d\}} |\mathcal{I}|
\end{align*}

As $\mathcal{P}$ contains all possible combinations of $W_t$'s columns and $R_t$'s rows for each $t \in \mathbf{d}$, it is not hard to see that $\forall \mathcal{S} \subset \bar{\mathbf{d}}$, $\forall t \in \mathbf{d}$, either all columns of $W_t$ appear in $w_{\mathcal{S}}$ or all rows of $R_t$ appear in $r_{\bar{\mathbf{d}} - \mathcal{S}}$. By Proposition \ref{prp:grk}, there exists $\mathcal{S}_0 \subset \bar{\mathbf{d}}$ such that $\rank w_{\mathcal{S}_0} + \rank r_{\bar{\mathbf{d}}-\mathcal{S}_0} = \min\limits_{\mathcal{S} \subset \bar{\mathbf{d}}} (\rank w_{\mathcal{S}} + \rank r_{\bar{\mathbf{d}}-\mathcal{S}})$. Now if all columns of $W_t$ appear in $w_{\mathcal{S}_0}$, remove rows of $R_t$ from $r_{\bar{\mathbf{d}}-\mathcal{S}_0}$ if any; else remove columns of $W_t$ from $w_{\mathcal{S}_0}$ if any. Repeat the removal process for all $t \in \mathbf{d}$, and denote the resultant matrices by $w_{\mathcal{S}_0}^-$ and $r_{\bar{\mathbf{d}}-\mathcal{S}_0}^-$ respectively. Then $\exists \mathcal{S}_1 \subset \mathbf{d}$ such that $w_{\mathcal{S}_0}^- = W_{\mathcal{S}_1}$ up to a permutation of columns and $r_{\bar{\mathbf{d}}-\mathcal{S}_0}^- = R_{\mathbf{d}-\mathcal{S}_1}$ up to a permutation of rows. So
\begin{align} \label{eqn:min-rank}
& \min_{\mathcal{S} \subset \bar{\mathbf{d}}} (\rank w_{\mathcal{S}} + \rank r_{\bar{\mathbf{d}}-\mathcal{S}}) \nonumber \\
\leq \enspace & \min_{\mathcal{S} \subset \mathbf{d}} \left( \rank W_{\mathcal{S}} + \rank R_{\mathbf{d} - \mathcal{S}} \right) \nonumber \\
\leq \enspace & \rank W_{\mathcal{S}_1} + \rank R_{\mathbf{d}-\mathcal{S}_1} \nonumber \\
= \enspace & \rank w_{\mathcal{S}_0}^- + \rank r_{\bar{\mathbf{d}}-\mathcal{S}_0}^- \nonumber \\
\leq \enspace & \rank w_{\mathcal{S}_0} + \rank r_{\bar{\mathbf{d}}-\mathcal{S}_0} \nonumber \\
= \enspace & \min_{\mathcal{S} \subset \bar{\mathbf{d}}} (\rank w_{\mathcal{S}} + \rank r_{\bar{\mathbf{d}}-\mathcal{S}})
\end{align}
It implies that 
\begin{align*}
& \grank \left( \sum_{i \in \mathbf{d}} W_i P_i R_i \right) 
= \grank \left( \sum_{i \in \bar{\mathbf{d}}} w_i p_i r_i \right) \\
= \enspace & \min_{\mathcal{S} \subset \bar{\mathbf{d}}} (\rank w_{\mathcal{S}} + \rank r_{\bar{\mathbf{d}}-\mathcal{S}}) \hspace{.3in} \text{By Proposition \ref{prp:grk}} \\
= \enspace & \min_{\mathcal{S} \subset \mathbf{d}} \left( \rank W_{\mathcal{S}} + \rank R_{\mathbf{d} - \mathcal{S}} \right) \hspace{.2in} \text{By (\ref{eqn:min-rank})}
\end{align*}
This completes the proof of Lemma \ref{lem:grk-mtrx}. \hfill $\qed$

Let $M \in \mathbb{C}^{n_1 \times n_2}$ be a complex matrix of rank $t \leq \min\{n_1, n_2\}$. By rank decomposition, there exist a full column rank matrix $W \in \mathbb{C}^{n_1 \times t}$ and a full row rank matrix $R \in \mathbb{C}^{t \times n_2}$ such that $M = WR$. In another word, $M = \sum\limits_{i=1}^t w_i r_i$, where $w_i \in \mathbb{C}^{n_1}$ is the $i$th column of $W$ and $r_i \in \mathbb{C}^{1 \times n_2}$ is the $i$th row of $R$. Let $\mathbf{t} \triangleq \{1, 2, \dots, t\}$. It is easy to see that $\{w_i \, | \, i \in \mathbf{t}\}$ and $\{r_i \, | \, i \in \mathbf{t}\}$ are both linearly independent sets.

\begin{lemma} \label{lem:const-lnr-rank}
Suppose a complex matrix $M_{n_1 \times n_2} = \sum\limits_{i=1}^t w_i r_i$, where $w_i \in \mathbb{C}^{n_1}$ are column vectors, $r_i \in \mathbb{C}^{1 \times n_2}$ are row vectors, $\{w_1, w_2, \dots, w_t\}$ and $\{r_1, r_2, \dots, r_t\}$ are both linearly independent sets. For some $d \geq t$, given column vectors $w_{t+1}$, $w_{t+2}$, $\dots$, $w_d \in \mathbb{C}^{n_1}$, row vectors $r_{t+1}$, $r_{t+2}$, $\dots$, $r_d \in \mathbb{C}^{1 \times n_2}$, and algebraically independent scalar parameters $p_1$, $p_2$, $\dots$, $p_d$, the following equation holds.
\begin{equation} \label{eqn:const-lnr-rank}
\grank \left( M + \sum_{i=t+1}^d w_i p_i r_i \right) = \grank \left( \sum_{i=1}^d w_i p_i r_i \right)
\end{equation}
\end{lemma}

\noindent \textbf{Proof of Lemma \ref{lem:const-lnr-rank}:} By the definition of generic rank, the left-hand side of equation (\ref{eqn:const-lnr-rank}) is less than or equal to the right-hand side, so it suffices to prove the other direction. There exists $\mathcal{S}_0 \subset \mathbf{d}$ such that $\mathcal{I}_0 = \{(w_i, r_i) \, | \, i \in \mathcal{S}_0\}$ is a jointly independent subset of $\{(w_i, r_i) \, | \, i \in \mathbf{d}\}$ with the maximum cardinality, i.e., $\max\limits_{\mathcal{I} \in \mathcal{J}\{w_i, r_i; d\}} |\mathcal{I}| = |\mathcal{I}_0|$.

First, it is claimed that equation (\ref{eqn:const-lnr-rank}) holds when $\mathbf{d} - \mathbf{t} \subset \mathcal{S}_0$. Now suppose the claim is true and consider the general case when $\mathbf{d} - \mathbf{t} \not\subset \mathcal{S}_0$. Let $\mathcal{S}_1 = \mathcal{S}_0 \cap (\mathbf{d} - \mathbf{t})$. We have
\begin{align*}
& \grank \left( \sum\limits_{i=1}^d w_i p_i r_i \right) \\
= \enspace & |\mathcal{I}_0| = \grank \left( \sum\limits_{i \in \mathcal{S}_0} w_i p_i r_i \right) &&\text{By Proposition \ref{prp:grk}} \\
= \enspace & \grank \left( \sum\limits_{i \in \mathbf{t} \cup \mathcal{S}_1} w_i p_i r_i \right) &&\text{By the definition of} \enspace \mathcal{I}_0 \\
= \enspace & \grank \left( M + \sum_{i \in \mathcal{S}_1} w_i p_i r_i \right) &&\text{By the claim} \\
\leq \enspace & \grank \left( M + \sum_{i=t+1}^d w_i p_i r_i \right) 
\end{align*}
So equation (\ref{eqn:const-lnr-rank}) holds for the general case.

Next, the claim will be proved. Now assume $\mathbf{d} - \mathbf{t} \subset \mathcal{S}_0$. Let $\mathbf{t}_1 = \mathcal{S}_0 \cap \mathbf{t}$. Let $\mathbf{t}_2 = \mathbf{t} - \mathbf{t}_1 = \mathbf{d} - \mathcal{S}_0$. As $\{w_i \, | \, i \in \mathbf{t}\}$ and $\{r_i \, | \, i \in \mathbf{t}\}$ are both linearly independent sets, and by the definition of $\mathcal{S}_0$, it is clear that each vector in $\{w_i \, | \, i \in \mathbf{t}_1\}$ (respectively, $\{r_i \, | \, i \in \mathbf{t}_1\}$) is linearly independent with the vectors in $\{w_i \, | \, i \in \mathbf{d} - \mathbf{t}_1\}$ (respectively, $\{r_i \, | \, i \in \mathbf{d} - \mathbf{t}_1\}$). So 
\begin{equation} \label{eqn:grk-part}
\grank \left( \sum\limits_{i=1}^d w_i p_i r_i \right) = |\mathbf{t}_1| + \grank \left( \sum\limits_{i \in \mathbf{d} - \mathbf{t}_1} w_i p_i r_i \right)
\end{equation}
Suppose the generic rank of $\left( \sum\limits_{i \in \mathbf{t}_2} w_i r_i + \sum\limits_{i=t+1}^d w_i p_i r_i \right)$ is admissible by $p_i = \bar{p}_i \in \mathbb{C}$ for $i \in \mathbf{d} - \mathbf{t}$, then
\begin{align*}
& \grank \left( M + \sum_{i=t+1}^d w_i p_i r_i \right) \\
\geq \enspace & \rank \left( \sum\limits_{i \in \mathbf{t}_1} w_i r_i + \sum\limits_{i \in \mathbf{t}_2} w_i r_i + \sum\limits_{i=t+1}^d w_i \bar{p}_i r_i \right) \\
= \enspace & \rank \left( \sum\limits_{i \in \mathbf{t}_1} w_i r_i \right) + \rank \left( \sum\limits_{i \in \mathbf{t}_2} w_i r_i + \sum\limits_{i=t+1}^d w_i \bar{p}_i r_i \right) \\
= \enspace & |\mathbf{t}_1| + \grank \left( \sum\limits_{i \in \mathbf{t}_2} w_i r_i + \sum\limits_{i=t+1}^d w_i p_i r_i \right)
\end{align*}
As $|\mathcal{S}_0| = |\mathbf{t}_1| + d - t$, by equation (\ref{eqn:grk-part}), $\grank \left( \sum\limits_{i \in \mathbf{d} - \mathbf{t}_1} w_i p_i r_i \right) = d - t$. Without loss of generality, assume $\mathbf{t}_1 = \{1, 2, \dots, t_1\}$. It is easy to see that 
\begin{align*}
& \mathcal{Q} \triangleq \left\lbrace \left( p_{t_1+1}, p_{t_1+2}, \dots, p_d \right) \in \mathbb{C}^{d-t_1} \, \Bigg| \, \right. \\
& \hspace{1.2in} \left. \rank \left( \sum\limits_{i \in \mathbf{d} - \mathbf{t}_1} w_i p_i r_i \right) = d - t \right\rbrace
\end{align*}
is an open set in $\mathbb{C}^{d-t_1}$, because its complement is a closed set. Let $p_{u,v} \in \mathbb{C}^{d-t_1}$ be the vector whose first $t-t_1$ entries are $u \in \mathbb{C}$ and last $d-t$ entries are $v \in \mathbb{C}$, i.e., $p_i = u$ for $i \in \mathbf{t}_2$ and $p_j = v$ for $j \in \mathbf{d} - \mathbf{t}$. Clearly $p_{0,1} \in \mathcal{Q}$, which implies that $\exists \delta > 0$ such that $p_{\delta,1} \in \mathcal{Q}$ thus $p_{1,\frac{1}{\delta}} \in \mathcal{Q}$. That is, $\grank \left( \sum\limits_{i \in \mathbf{t}_2} w_i r_i + \sum\limits_{i=t+1}^d w_i p_i r_i \right) \geq d - t$. Therefore, 
\begin{align*}
& \grank \left( M + \sum_{i=t+1}^d w_i p_i r_i \right) \\ 
\geq \enspace & |\mathbf{t}_1| + d - t = |\mathcal{S}_0| \\
= \enspace & \grank \left( \sum_{i=1}^d w_i p_i r_i \right)
\end{align*}
The claim is proved. This completes the proof. \hfill $\qed$

\begin{corollary} \label{cor:const-lnr-mtrx-rank}
Suppose a complex matrix $M_{n_1 \times n_2} = \sum\limits_{i=1}^t w_i r_i$, where $w_i \in \mathbb{C}^{n_1}$ are column vectors, $r_i \in \mathbb{C}^{1 \times n_2}$ are row vectors, $\{w_1, w_2, \dots, w_t\}$ and $\{r_1, r_2, \dots, r_t\}$ are both linearly independent sets. For some $d \geq t$, given complex matrices $W_{t+1}$, $W_{t+2}$, $\dots$, $W_d$, each of which has $n_1$ rows and an arbitrary number of columns, complex matrices $R_{t+1}$, $R_{t+2}$, $\dots$, $R_d$, each of which has an arbitrary number of rows and $n_2$ columns, algebraically independent scalar parameters $p_1$, $p_2$, $\dots$, $p_t$, and parameterized matrices $P_{t+1}$, $P_{t+2}$, $\dots$, $P_d$ of appropriate sizes, whose entries are algebraically independent thus are modeled by distinct parameters, the following equation holds.
\begin{align*}
& \grank \left( M + \sum_{i=t+1}^d W_i P_i R_i \right) \\
= \enspace & \grank \left( \sum_{i=1}^t w_i p_i r_i + \sum_{i=t+1}^d W_i P_i R_i \right)
\end{align*}
\end{corollary}

\noindent \textbf{Proof of Corollary \ref{cor:const-lnr-mtrx-rank}:} For each $s \in \mathbf{d} - \mathbf{t}$, suppose $W_s$ is of size $n_1 \times \alpha_s$, $R_s$ is of size $\beta_s \times n_2$, and $P_s$ is of size $\alpha_s \times \beta_s$. Then
\begin{align*}
& \grank \left( M + \sum_{s=t+1}^d W_s P_s R_s \right) \\
= \enspace & \grank \left( M + \sum_{s=t+1}^d \sum_{i=1}^{\alpha_s} \sum_{j=1}^{\beta_s} w_s^i p_s^{ij} r_s^j \right) \\
= \enspace & \grank \left( \sum_{s=1}^t w_s p_s r_s + \sum_{s=t+1}^d \sum_{i=1}^{\alpha_s} \sum_{j=1}^{\beta_s} w_s^i p_s^{ij} r_s^j \right) \\
= \enspace & \grank \left( \sum_{s=1}^t w_s p_s r_s + \sum_{s=t+1}^d W_s P_s R_s \right)
\end{align*}
where $w_s^i \in \mathbb{C}^{n_1}$ is the $i$th column of $W_s$, $p_s^{ij}$ is the $ij$th entry of $P_s$, and $r_s^j \in \mathbb{C}^{1 \times n_2}$ is the $j$th row of $R_s$. \hfill $\qed$

\noindent \textbf{Proof of Proposition \ref{prp:mtrx-pencil}:} (Sufficiency) If $\exists \mathcal{S} \subset \mathbf{k}$ such that (\ref{eqn:real-partition}) holds, by Lemma \ref{lem:pencil-rank},  
\begin{equation*}
\rank \left( \lambda I_n - A - B_{\mathcal{S}} E_{\mathcal{S}} - E_{\mathbf{k}-\mathcal{S}} C_{\mathbf{k}-\mathcal{S}} \right) < n
\end{equation*}
for any matrices $E_{\mathcal{S}} \in \mathbb{R}^{m_{\mathcal{S}} \times n}$ and $E_{\mathbf{k}-\mathcal{S}} \in \mathbb{R}^{n \times l_{\mathbf{k}-\mathcal{S}}}$, where $m_{\mathcal{S}} \triangleq \sum\limits_{i \in \mathcal{S}} m_i$ and $l_{\mathbf{k}-\mathcal{S}} \triangleq \sum\limits_{j \in \mathbf{k}-\mathcal{S}} l_j$. It implies that 
\begin{equation*}
\rank \left( \lambda I_n - A - \sum_{i \in \mathbf{k}} B_i F_i C_i \right) < n
\end{equation*}
for any matrices $F_i \in \mathbb{R}^{m_i \times l_i}$, $i \in \mathbf{k}$. Thus $\lambda$ is in the fixed spectrum of system $\{A, B_i, C_i; k\}$.

(Necessity) If $\lambda$ is in the fixed spectrum of system $\{A, B_i, C_i; k\}$, 
\begin{equation*}
\rank \left( \lambda I_n - A - \sum_{i \in \mathbf{k}} B_i F_i C_i \right) < n
\end{equation*}
for any matrices $F_i \in \mathbb{R}^{m_i \times l_i}$, $i \in \mathbf{k}$. Suppose $\rank (A - \lambda I_n) = t < n$, then there exist $w_i \in \mathbb{C}^n$ and $r_i \in \mathbb{C}^{1 \times n}$, $i \in \mathbf{t}$, such that $\{w_i \, | \, i \in \mathbf{t}\}$ and $\{r_i \, | \, i \in \mathbf{t}\}$ are both linearly independent sets, and $A - \lambda I_n = \sum\limits_{i=1}^t w_i r_i$. By Corollary \ref{cor:const-lnr-mtrx-rank}, 
\begin{equation*}
\grank \left( \sum_{i=1}^t w_i p_i r_i + \sum_{j=1}^k B_j F_j C_j \right) < n
\end{equation*}
By Lemma \ref{lem:grk-mtrx}, $\exists \mathcal{S}_1 \subset \mathbf{t}$, $\mathcal{S}_2 \subset \mathbf{k}$ such that 
\begin{equation*}
\rank [w_{\mathcal{S}_1} \enspace B_{\mathcal{S}_2}] + \rank 
\begin{bmatrix}
r_{\mathbf{t} - \mathcal{S}_1} \\
C_{\mathbf{k} - \mathcal{S}_2}
\end{bmatrix}
< n
\end{equation*}
So 
\begin{align*}
& \rank [w_{\mathcal{S}_1} \enspace B_{\mathcal{S}_2}] 
\begin{bmatrix}
r_{\mathcal{S}_1} \\
E_{\mathcal{S}_2}
\end{bmatrix}
+ 
\rank [w_{\mathbf{t} - \mathcal{S}_1} \enspace E_{\mathbf{k} - \mathcal{S}_2}]
\begin{bmatrix}
r_{\mathbf{t} - \mathcal{S}_1} \\
C_{\mathbf{k} - \mathcal{S}_2}
\end{bmatrix} \\
& \hspace{3in} < n
\end{align*}
for any matrices $E_{\mathcal{S}_2}$ and $E_{\mathbf{k} - \mathcal{S}_2}$ of appropriate sizes. As $A - \lambda I_n = w_{\mathcal{S}_1} r_{\mathcal{S}_1} + w_{\mathbf{t} - \mathcal{S}_1} r_{\mathbf{t} - \mathcal{S}_1}$, it implies that
\begin{equation*}
\rank \left( A - \lambda I_n + B_{\mathcal{S}_2} E_{\mathcal{S}_2} + E_{\mathbf{k} - \mathcal{S}_2} C_{\mathbf{k} - \mathcal{S}_2} \right) < n
\end{equation*}
for any matrices $E_{\mathcal{S}_2}$ and $E_{\mathbf{k} - \mathcal{S}_2}$. By Lemma \ref{lem:pencil-rank}, (\ref{eqn:real-partition}) holds for $\mathcal{S}_2 \subset \mathbf{k}$. \hfill $\qed$

\section{Concluding Remarks}

This paper utilizes matroid theory to simplify the proof of the algebraic condition derived in \cite{anderson1981algebraic} for the existence of a fixed spectrum in a multi-channel linear system.

\section*{Acknowledgement}

The authors would like to thank Brian D. O. Anderson for suggesting the problem which this paper is addressed.



\bibliographystyle{IEEEtran}
\bibliography{fjbib-fix-spec}


\end{document}